\documentclass[
preprint,
 amsmath,amssymb,
 aps,
prb,
]{revtex4-1}

\usepackage{graphicx}
\usepackage{physics}

\newcommand{\rmi}{\mathrm{i}}
\newcommand{\rme}{\mathrm{e}}
\newcommand{\vect}[1]{\mathbf{#1}}
\newcommand{\rr}{\vect{r}}

\newcommand{\boga}{\boldsymbol{\gamma}}
\newcommand{\boal}{\boldsymbol{\alpha}}
\newcommand{\bosi}{\boldsymbol{\sigma}}

\begin{document}

\title{Weyl fermions in cylindrical wires}

\author{Mikhail Erementchouk}%
 \email{merement@umich.edu}
 \author{Pinaki Mazumder}%
 \email{pinakimazum@gmail.com}
\affiliation{%
 Department of Electrical Engineering and Computer Science,\ 
 University of Michigan, Ann Arbor, MI 48109 USA 
}%

\begin{abstract}
The key feature of Weyl semimetals (WSM) is the presence of topologically protected Dirac cones in a 3D material. 
We consider the effect of restricting geometry on the spectrum of excitations in WSM using as a model a cylindrical WSM wire. For the full manifold of hard boundary conditions, we derive the general form of the dispersion equation relating the energy of the excitations and their momentum along the wire. We show that only the special class of boundary conditions, corresponding to decoupled helicities or, equivalently, to pinned directions of the electron spin on the surface, support massless excitations. For a general boundary condition, these excitations acquire mass inversely proportional to the radius of the wire. This demonstrates that boundary phenomena may play a crucial role in formation of excitations in WSM based structures.
\end{abstract}

\maketitle

\section{Introduction}
\label{sec:intro}

Recent synthesis of Weyl semimetals (WSM)\cite{yang_weyl_2015,xu_observation_2015,xu_discovery_2015,lv_observation_2015,lv_experimental_2015,lu_experimental_2015} marks a significant achievement in a series of efforts invested in finding solid state implementations of Weyl materials, whose unique properties were envisioned almost a century ago \cite{weyl_elektron_1929}. The principal feature of the new emergent class of materials such as $\mathrm{TaAs} $, $\mathrm{NbAs} $ is that they realize WSM in stoichiometric single crystals rather than in a material with carefully crafted chemical composition as, for instance, in $Q-\mathrm{Bi}_{1-x} \mathrm{Sb}_x \mathrm{Te}_3$ with $Q=\mathrm{La}$ or $\mathrm{Lu}$, where the WSM phase was expected for $0.39<x<0.42$ and $0.41<x<0.46$, respectively\cite{prev_1_Liu}, or in a complex heterostructure as, for example, in $\mathrm{HgTe}/\mathrm{CdTe}$ multilayers\cite{prev_2_Halasz}. Owing to the relative simplicity of the discovered materials, WSM started to leave the realm of the theoretical high-energy physics and to attract significant attention from more ``everyday'' perspectives including possible applications of WSM in a new type of electronics, weyltronics, based on unique features of WSM \cite{shekhar_extremely_2015,wang_helicity-protected_2016}. The property that sets WSM apart from other Dirac materials is the separation of Dirac points corresponding to states with different helicities, which provides means for dynamical distinction between such states. This property of WSM is a subject of an intensive of fundamental \cite{landsteiner_anomalous_2014,landsteiner_notes_2016,gorbar_second-order_2017,qiu_electrodynamics_2017,cortijo_strain-induced_2016,grushin_inhomogeneous_2016,gorbar_consistent_2017,zubkov_emergent_2015,hidaka_relativistic_2017} and more application oriented \cite{hosur_recent_2013,bai_wavevector_2016,bai_chiral_2016,erementchouk_helicity-dependent_2017} research.

At the same time, the main attention is paid to either infinite systems or to structures with simple flat surfaces thus leaving open the question of the effect of restricting geometry. The general objective of the present paper is to address this question by discussing main spectral features of excitations propagating in a WSM wire of finite radius.

One of the difficulties in describing finite WSM structures is to specify correctly the hard boundary conditions, which ensure that the electrons remain inside the material. The problem of boundary conditions imposed on solutions of the Dirac equation started to attract special attention in the context of states with reduced dimensionality (see Ref.~\onlinecite{volkov_surface_2016} and references therein) almost half-century ago but still is a subject of research \cite{devizorova_fermi_2017,li_spiraling_2015,hashimoto_boundary_2017}. One of the reasons for such delayed development is a drastic difference between the dynamical origins of the hard boundary for the canonical Schrodinger case, which essentially follows prescriptions of the classical mathematical physics, and for a Dirac particle. In the non-relativistic case, the hard boundary is equivalent to the presence of a sufficiently high potential barrier. Such barrier supports only attenuated solutions in the prohibited region, which eventually leads to the simple requirement of vanishing Schrodinger wave function at the boundary. This approach, however, doesn't work for a Dirac equation due to the Klein tunneling \cite{katsnelson_chiral_2006,tudorovskiy_chiral_2012,reijnders_semiclassical_2013,kleptsyn_chiral_2015}. Indeed, the scalar potential raises the level of the Dirac sea and, as a result, a sufficiently high scalar potential barrier instead of blocking propagation opens new propagating channels. Thus, a scalar potential cannot support the hard boundary and a more general class of matrix potentials \cite{mccann_symmetry_2004,akhmerov_detection_2007,akhmerov_boundary_2008} must be considered yielding a manifold of hard boundary conditions.

We show that despite the fact that confining surfaces may be described by various boundary conditions, it is possible to outline general spectral features, which hold at almost all boundary conditions. One of such common features is a series of bands of massive excitations existing at positive and negative energies. The existence of such bands may appear a natural consequence of hard boundary conditions similarly to those in classical electrodynamic and quantum Schrodinger systems. It is well known, however, that this picture does not hold of systems governed by the Dirac equation, where, for instance, near hard boundaries may exist surface, or edge, states, which are impossible in the canonical systems. 
We show that, in WSM wires with the finite radius, such additional states are in general massive due to the boundary induced coupling between states with different helicities. The only class of boundary conditions admitting massless excitations is the one corresponding to fully decoupled helicities and, as a consequence, prescribing the definite orientation of the electron spin at the surface of the wire.

\section{Dispersion equations of guided modes}

Within the model with two Weyl points, the dynamics of Weyl fermions is described by a Dirac equation with a helicity dependent scalar and vector potentials accounting for separation of the Weyl points
\begin{equation}\label{eq:full_Dirac}
 \rmi \left( \gamma^0 \partial_0 + v \boga\cdot \nabla \right) \widetilde{\Psi} = \widehat{U}(\rr) \widetilde{\Psi},
\end{equation}
where we have introduced the full potential
\begin{equation}\label{eq:full_potential}
 \widehat{U}(\rr) = e V(\rr)  + \gamma^5 q_0(\rr) +
 			v \boga \cdot \left[e \vect{A}(\rr) + \gamma^5 \vect{q}(\rr) \right].
\end{equation}
Here, $V(\rr)$ and $\vect{A}(\rr) $ are the scalar and vector potential, respectively, and $q_0(\rr) $ and $\vect{q}(\rr) $ are half-distances between the Weyl points in the energy and in the momentum space.

We employ the fact that both physical potentials and the separations between the Weyl points are spatially uniform inside the wire and thus vector potentials can be excluded with the help of a gauge transformation
\begin{equation}\label{eq:gauge_transformation}
\widetilde{\Psi}(\rr)  = \exp \left\{ - \rmi e \vect{A} \cdot \rr - \rmi \gamma^5 \vect{q} \cdot \rr \right\}
 \Psi (\rr).
\end{equation}
Additionally, taking into account that spatially uniform physical potentials lead to simple shifts of the energy and the momentum, we assume that $V = 0$ and $\vect{A} =0 $.
This turns~\eqref{eq:full_Dirac} into an equation with $\widehat{U} = 0$ and thus possessing the cylindrical symmetry. Furthermore, we make use of the translational symmetry along the axis of the wire and separate longitudinal and transversal variables. Thus, distinguishing sectors with opposite helicities 
\begin{equation}\label{eq:psi_helicities}
 \Psi(\rr) = \mqty(\psi_+ \\ \psi_-),
\end{equation}
and choosing the $z$-axis along the axis of the wire, we obtain the equations of motion within each sector
\begin{equation}\label{eq:separated_Dirac}
 \left( \epsilon_\xi + \xi k_\xi \sigma_z \right) \psi_\xi(\rr_\perp) = \rmi \xi \bosi_\perp \cdot \vect \nabla_\perp \psi_\xi(\rr_\perp),
\end{equation}
where $\epsilon_\xi = \epsilon - \xi q_0 $,  $k_\xi = k + \xi q_z $, $\rr_\perp$ is the radius-vector in the $(x, y) $-plane, $\bosi_\perp \cdot \vect \nabla_\perp = \sigma_x \partial/\partial x + \sigma_y \partial/\partial y$, and we have chosen units with $v = 1$.

Finally, the rotational symmetry yields the representation
\begin{equation}\label{eq:psi_rotational}
 \psi_\xi(\rr_\perp) = \sum_{m = -\infty}^{\infty} \rme^{\rmi (m +1/2) \phi} \widehat{R}_z (\phi)
 			\psi^{(j)}_\xi(r),
\end{equation}
where $\widehat{R}_z (\phi) =  \rme^{-\rmi \sigma_z \phi/2}$ accounts for rotation of the spin while encircling the origin and $j = m + 1/2$ is the $z$-projection of the total angular momentum.

A general solution of the radial equation corresponding to projection $j$, up to a normalization factor, can be written as
\begin{equation}\label{eq:psi_radial}
 \psi^{(j)}_\xi(r) = \mqty(f_\xi^{(1)} J_{j-1/2}(Q_\xi r) \\
 						\rmi f_\xi^{(2)} J_{j+1/2}(Q_\xi r)),
\end{equation}
where $J_n$ are the Bessel functions of the first kind, $Q_\xi = \sqrt{\epsilon_\xi^2 - k_\xi^2} $, and we have chosen $f_\xi^{(1,2)} = \xi k_\xi \mp (\xi Q_\xi + \epsilon_\xi) $ in the form emphasizing a symmetry between the components with positive and negative $j$. In the latter case, one needs to use the  relation $J_{-|n|} = (-1)^n J_{|n|}$. 

\begin{figure}[tb]
	\centering
	\includegraphics[width = 3in]{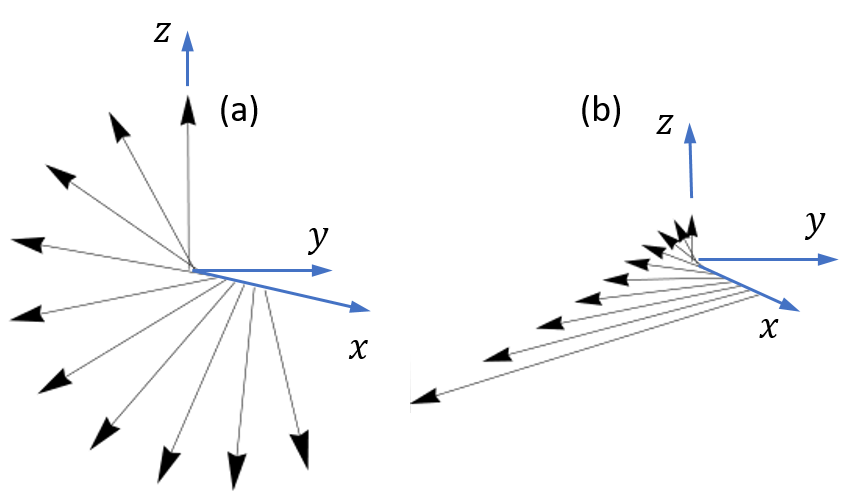}
	\caption{The variation of the spin in a state propagating in a WSM wire with the axis parallel to the $z$-axis along the radius (the $x$-axis) in (a) extended and (b) surface modes. At the origin, the spin is parallel or anti-parallel to the $z$-axis for $j  = \pm 1/2$. The spin is confined to the $(y,z)$-plane for all $x$ and experiences full revolutions (for sufficiently large $r$) for $Q_\xi^2 > 0$ and hyperbolic-like rotation for $Q_\xi^2 < 0$. }
	\label{fig:spin_state}
\end{figure}

Depending on whether $Q_\xi$ is real ($\epsilon_\xi^2 > k_\xi^2 $) or imaginary ($\epsilon_\xi^2 < k_\xi^2 $), the state described by Eq.~\eqref{eq:psi_radial} either extends over the whole cross-section of the wire or is localized near the boundary thus forming  surface (edge) states. The case $Q_\xi = 0$ corresponds to an algebraic variation with the distance to the axis of the wire and requires a special consideration (see Appendix~\ref{sec:zero_mass_app}). An important feature of the radial dependence of the spin state must be emphasized: at the axis of the wire, the spin is oriented along or against the axis depending on the sign of the angular momentum and, away from the axis of the wire, the spin lies in the plane normal to the radius (see Fig.~\ref{fig:spin_state}). 

% Solutions~\eqref{eq:psi_radial} are subjects to boundary conditions. WSM systems with restriced geometry admit a variety of boundary conditions, whose discussion we postpone to the next section. Here, we discuss general properties of modes described by Eq.~\eqref{eq:psi_radial}, which hold for any particular choice of the boundary conditions.

% ... For propagating modes ($k_\xi^2 > 0$), within each sector with definite helicity, there is no radial component of the current
% \begin{equation}\label{eq:radial_current_xi}
%   j_r \equiv \matrixel{\psi_\xi^{(j)}}{\sigma_x}{\psi_\xi^{(j)}} = 0.
% \end{equation}

% ... Thus, the current is tangent to the cylindrical surface

% ... Typical form of the current distribution is shown in Fig. A(a) (bulk state) and Fig. A(b) (edge state)

% \tobedone{Pretty pictures with the current distribution}

% ... Coexistence of different forms of spatial distributions for different helicities (this also does not depend on helicity since the longitudinal wavenumber is a good quantum number)

% \section{Dispersion equations}

Solutions~\eqref{eq:psi_radial} are subjects to boundary conditions. As has been discussed in Introduction, there is a manifold of possible boundary conditions corresponding to fermions confined to the interior of the wire. Restrictions imposed on this manifold translate into physical conditions, which, while may take place in particular systems, may be too restrictive to be adopted on the general ground. For example, requiring decoupled character of states with opposite helicities implies rather special requirement for the spin of Weyl fermions to have definite orientation on the surface of the wire. Indeed, absent radial current, $ \left. \matrixel{\Psi}{\boal\cdot\vect{n}_B}{\Psi} \right|_{r = r_B}= 0$, means that within each helicity sector the spin should be tangent to the surface of the wire and then, due to linearity, all states must have spin with the same orientation. 
% Conversely, completely relaxed requirement with regard the spin state on the surface of the wire can only take place in systems with a strong relation between states with opposite helicities at the surface. 

%Boundary conditions of a general form should admit all intermediate situations. 

As we show in Appendix~\ref{sec:bc_app}, a general boundary condition imposed at the surface of the wire can be written as
\begin{equation}\label{eq:Psi_BC}
 \widehat{M} \Psi(r_B) = \Psi(r_B),
\end{equation}
with the matrix $\widehat{M} $ having in the rotated frame, $\widetilde{M} = \rme^{ \rmi \Sigma_z \phi/2} \widehat{M} \rme^{ -\rmi \Sigma_z \phi/2}$, the form
\begin{equation}\label{eq:M_paramterized}
 \widetilde{M} = \widehat{M}_\|(\vect{v}_+, \vect{v}_- ) \cos{\Lambda}
 			+ \widehat{M}_\perp(\theta, \chi ) \sin{\Lambda},
\end{equation}
where $\Lambda $ is the helicity mixing angle, unit vectors $\vect{v}_\xi $ lying in the $(y,z)$-plane describe the ``pinned'' spin states of Weyl fermions with helicity $\xi$, and the matching angle $\theta$ quantifies the rotation around the $x$-axis aligning vectors $\vect{v}_\xi $ so that $\vect{v}_+ = - R_x(\theta)\vect{v}_- $.

% \tobedone{Alternatively, this can be interpreted in the form of the variety of boundary conditions available in the two-valley setup: we either have pinned states or coupled valleys (or everything in between)}

Imposing condition~\eqref{eq:Psi_BC} on a solution, one finds a dispersion equation $D(\epsilon, k) = 0$ relating the energy and the longitudinal momentum of the modes propagating along the wire with
\begin{equation}\label{eq:D_full}
 D(\epsilon, k) = \frac{1}{2}
 		\matrixel{\overline{\psi}_-}{
 			R_x^{-1}(\theta) \left( \sigma_0 - \vect{v}_+ \cdot \bosi \cos(\Lambda) \right)}{\psi_+},
\end{equation}
where $\bra{\overline{\psi}_-} $ is a state satisfying 
\begin{equation}\label{eq:bar_psi_def}
 \braket{\overline{\psi}_-}{- \vect{u}} = -\braket{\vect{u}}{\psi_-}
\end{equation}
for any unit vector $\vect{u}$ in the $(y,z) $-plane, so that for $\psi_- = \left( \psi_-^{(1)}, \psi_-^{(2)} \right)^T $ one has $\bra{\overline{\psi}_-}  = \rmi \left( \psi_-^{(2)}, -\psi_-^{(1)} \right)$.

In the limit $\Lambda = 0$, Eq.~\eqref{eq:D_full} yields the dispersion equation for decoupled helicities with the spin states related by $\vect{v}_+ = - R_x(\theta)\vect{v}_- $, as discussed in Appendix~\ref{sec:bc_app}. In this case, the dispersion equation factorizes $D(\epsilon, k) =  D_+(\epsilon, k; \vect{v}_+) D_-(\epsilon, k; \vect{v}_-) $, where
\begin{equation}\label{eq:disp_decoupled}
D_\pm(\epsilon, k; \vect{v}_+) = \braket{-\vect{v}_\pm}{\psi_\pm}.
\end{equation}

In the opposite limit of strong coupling, $\Lambda = \pi/2$, the dispersion equation ensures that the spin states of fermions with different helicities are directly related to each other at the surface of the wire. In this case, one has $D(\epsilon, k) = D(\epsilon, k; \theta) $, with
\begin{equation}\label{eq:D_strong_coupling}
 D(\epsilon, k; \theta) = \matrixel{\overline{\psi}_-}{R_x^{-1}(\theta) }{\psi_+}.
\end{equation}

Generally, expanding $\sigma_0$ in~\eqref{eq:D_full} in terms of eigenstates of $\vect{v}_+ \cdot \bosi$, the dispersion equation can be presented in the form explicitly showing the transition between these limiting cases
\begin{equation}\label{eq:D_full_sep}
\begin{split}
D (\epsilon, k) = & D(\epsilon, k; \theta) \sin^2(\Lambda/2) \\
	&- 
  		D_+(\epsilon, k; \vect{v}_+) D_-(\epsilon, k; \vect{v}_-) \cos(\Lambda).
\end{split}
\end{equation}

% The derived dispersion equations are one of the main results of the paper. While they were obtained  assuming the cylindrical symmetry, the most general form of the dispersion equation can be found by transforming back to the stationary frame. For example, instead of~\eqref{eq:D_full}, we have
% \begin{equation}\label{eq:D_full_gen}
% \begin{split}
%  D(\epsilon, k; \rr_B) & = \frac{1}{2}
%  		\bra{\overline{\psi}_-(\rr_B)}
%  			R_B^{-1}(\theta(\rr_B)) \\
% %
% 	& \times	\left[ \sigma_0 - \vect{v}_+(\rr_B) \cdot \bosi \cos(\Lambda(\rr_B)) \right] \ket{
% 			\psi_+(\rr_B)},
% \end{split}
% \end{equation}
% where $\rr_B$ is a point at the surface and $R_B^{-1}(\theta(\rr_B)) $ is the rotation operator around the outward normal to the surface at that point. Equations $D(\epsilon, k; \rr_B) = 0$ must hold for all $\rr_B$ and thus can be used for studying systems where the cylindrical symmetry is broken either by $\vect{A} $ and $\vect{q} $ having components in the plane orthogonal to the axis of the wire or by directionally dependent boundary conditions.

\section{Spectra of propagating states}

We limit our analysis of the dispersion relations to the case when the system possesses the full cylindrical symmetry. This takes place when $\vect{q}_\perp \equiv 0 $, so that the Weyl points may only be separated in energy and along the $z$-axis in the momentum space. % In this case, the spectrum of propagating modes is that of states with definite $z$-component of the total angular momentum, $D(\epsilon, k) = \prod_j D^{(j)}(\epsilon, k) $, where $D^{(j)}(\epsilon, k) $ are given by~\eqref{eq:D_full_sep} with $\psi_\xi = \psi^{(j)}_\xi $ and $\bra{\overline{\psi}_-} = -\left(f_-^{(2)} J_{j+1/2}(Q_\xi r), \rmi f_-^{(1)} J_{j-1/2}(Q_\xi r) \right) $. 
Even in this case, the spectrum of fermions in WSM wires is very feature-rich owing to multidimensionality of the manifold of boundary conditions. We, therefore, limit ourselves to discussing general spectral properties, which hold for a wide variety of boundary conditions, and pay the most attention to an important effect of the restricting geometry of WSM wires: boundary induced mass of excitations.

\subsection{Decoupled helicities}
\label{sec:decoupled}

In the case of decoupled helicities, the effect of the separation between the Weyl points reduces to simple shifts of the energy and the longitudinal wavenumber. Therefore, in order to simplify the notations, we assume that the Weyl points are not separated and omit the helicity index, $\xi$, where it is irrelevant.

Making use of the explicit expressions for $\psi^{(j)}_\xi $, we obtain
\begin{equation}\label{eq:De_0}
\begin{split}
 D_\xi(\epsilon, k; \vect{v}_\xi) = &\sin(\beta_\xi/2) f_\xi^{(1)} J_{j-1/2}(Q r_B) \\
 			&	+ \cos(\beta_\xi/2) f_\xi^{(2)} J_{j+1/2}(Q r_B),
\end{split}
\end{equation}
where $\beta_\xi$ is the angle between $\vect{v}_\xi $ and the $z$-axis.

It must be noted that $D_\xi(\epsilon, k; \vect{v}_\xi)$ identically vanishes at $\epsilon = \xi k $. This is the consequence of inadequate representation of solutions for the radial equation in form~\eqref{eq:psi_radial} for the case when $Q = 0$. We present a detailed analysis of this case in Appendix~\ref{sec:zero_mass_app}, where we show that modes corresponding to $\epsilon = \xi k $ may exist only for boundary conditions of the special form, $\vect{v}_\xi = - \xi \vect{e}_z $. In what follows, we assume that this condition is not met and, therefore, solutions of $D_\xi(\epsilon, k; \vect{v}_\xi) = 0$ corresponding to $\epsilon = \xi k $ must be excluded.

The gapped solutions of $D_\xi(\epsilon, k; \vect{v}_\xi) = 0 $ correspond to standing cylindrical waves inside the wire and thus emerge at energies yielding relatively large values of arguments of the Bessel functions in~\eqref{eq:De_0}. Taking this into account, we obtain in the limit of small $k$ the dispersion laws of the massive states for $\epsilon > 0$ and $j = 1/2$ in the form
 \begin{equation}\label{eq:massive_positive}
  \epsilon^{(l)}_\xi(k) = \mu_\xi^{(l)} + \frac{k^2}{2\mu_\xi^{(l)}},
 \end{equation}
 where integer $l$ enumerates subbands, and the masses of the excitations are
 \begin{equation}\label{eq:mass_positive}
  \mu^{(l)}_\xi = \frac{1}{2 r_B} \left( \beta_\xi + \frac{\pi}{2} + 2 \pi l \right).
 \end{equation}
For negative energies, the subbands have negative masses $\epsilon_\xi^{(l)}(k) = -\mu_\xi^{(l)} - \frac{k^2}{2\mu_\xi^{(l)}}$ with $\mu_\xi^{(l)} = \left( \beta_\xi + 2 \pi l - \pi/2 \right)/2r_B$. 

The presence of massive excitations may appear a natural consequence of the hard boundary conditions (cf. a Schrodinger particle in a confining potential or the electromagnetic field in a conducting hollow waveguide). It is important, therefore, to emphasize that, despite strong confinement, in addition to massive states there are also massless excitations. In order to qualitatively describe them, we consider the solutions of $D_\xi(\epsilon, k; \vect{v}_\xi) = 0$ near the point where they cross the $Q = 0$ states with a non-trivial spin orientation (see Appendix~\ref{sec:zero_mass_app}). Employing the smallness of $Q$, we can approximate Eq.~\eqref{eq:psi_radial} by
\begin{equation}\label{eq:psi_appr}
 \psi_\xi^{(j)} \propto \mqty(-1 \\ (\xi \epsilon_\xi + k_\xi)\frac{ \rmi }{K^{(j)}}), 
\end{equation}
where 
\begin{equation}\label{eq:K_def}
 K^{(j)} = \frac{2j +1}{r_B}
\end{equation}
defines a characteristic spatial and energy scale induced by the finite radius of the wire.

Using Eq.~\eqref{eq:psi_appr}, we find the spectrum of massless excitations for positive projections of the total angular momentum
\begin{equation}\label{eq:pinned_massless}
 \epsilon^{(j)}_\xi(k) = K^{(j)}(\beta_\xi) - \xi k,
\end{equation}
where $K^{(j)}(\beta_\xi) = K^{(j)} \tan(\beta_\xi/2) $. 

For $j < 0$, we have $\epsilon^{(j)}_\xi(k) = - \left[ \bar{K}^{(j)}(\beta_\xi) - \xi k \right]$, with $\bar{K}^{(j)}(\beta_\xi) = (2|j|+1) r_B^{-1} \cot(\beta_\xi/2) $.

\begin{figure}[tb]
	\centering
	\includegraphics[width=3in]{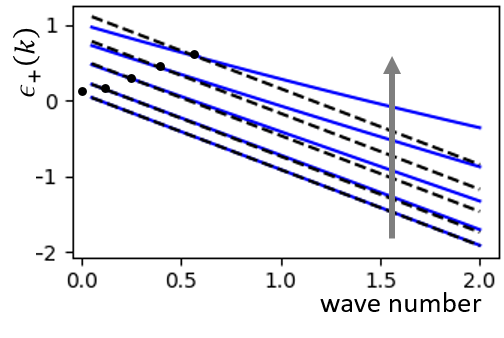}
	\caption{The dispersion laws, $\epsilon_+(k)$, of the gapless modes obtained numerically (solid lines) and from~\eqref{eq:pinned_massless} (dashed lines) for different angles between the pinned spin state and the $z$-axis, $\beta_+ = 5^{\circ}, 15^{\circ}, 30^{\circ}, 45^{\circ}, 60^{\circ}$. The arrow shows the variation of the lines with increasing $\beta_+$. The dots indicate the transition from the state extending over the cross-section of the wire ($k < K^{(j)}(\beta_+)/2 $) to surface states ($k > K^{(j)}(\beta_+)/2 $).}
	\label{fig:pinned_gapless}
\end{figure}

Figure~\ref{fig:pinned_gapless} presents the results of the numerical solution of $D_+(\epsilon, k; \vect{v}_+) = 0 $ together with the results obtained from~\eqref{eq:pinned_massless} for different orientations of the pinned spin state. It shows that for small $\beta_\xi$, when $K^{(j)}(\beta_\xi) r_B $ is small, approximation~\eqref{eq:psi_appr} reproduces main spectral features satisfactorily.

The massless states change their character from extending over the cross-section of the wire (for $|k| < K^{(j)}(\beta_\xi)/2 $) to surface modes (when $|k| > K^{(j)}(\beta_\xi)/2 $). Thus, in this case, the dispersion curves of the surface states are rays with the termination points $k_c = K^{(j)}(\beta_\xi)/2 $ except for $\beta_\xi = 0$, when the whole spectral branch corresponds to the spin distribution algebraically decaying with the distance from the surface of the wire.

\subsection{Strong coupling between helicities}
\label{sec:strong_coupling}

A similar analysis as in the previous section can be used for discussion of some general spectral features in the case of strong coupling, $\Lambda = \pi/2$. Indeed, Eq.~\eqref{eq:D_strong_coupling} can be interpreted as if the pinned state at the surface of the wire is given by the spin state of the fermions with the complementary helicity. Since the general form of the features discussed above does not depend on the precise orientation of the boundary spin, they preserve in the limit of strong coupling as well.

The main changes induced by the strong coupling happen at a vicinity of anti-crossing of dispersion curves corresponding to decoupled helicities. Near these points, we can use an approximation similar to~\eqref{eq:psi_appr} for both $\ket{\psi_+} $ and $\bra{\bar{\psi}_-} $. Expanding the resultant equation, one can see that it contains the energy, the longitudinal momentum and the separation between the Weyl points only in combinations
\begin{equation}\label{eq:ek_hom_pars}
 \bar{\epsilon} = \epsilon + q_z, \qquad 
 		\bar{k} = k - q_0.
\end{equation}
Thus, the spectrum of systems with different separations will have the same form as discussed below up to shifts in the $(\epsilon, k) $-plane. It should be noted in this regard that the separation of the Weyl points in energy, $q_0$, induces shifts along the $k$-axis, while $q_z$ results in shifts along the energy axis.

\begin{figure}[tb]
	\centering
	\includegraphics[width=3in]{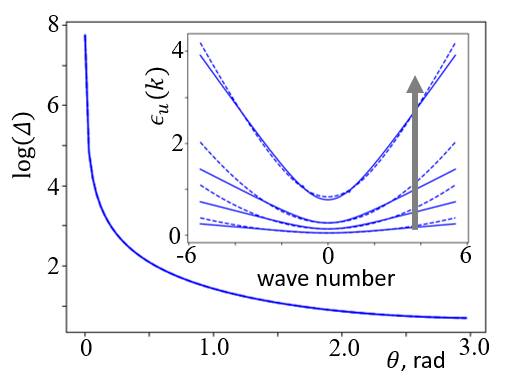}
	\caption{Natural logarithm of the half-width of the gap (or the mass) induced by the boundary condition corresponding to a strong coupling as a function of the matching angle $\theta$. The numerical and analytical results are indistinguishable on this scale. The inset shows the upper branch of the massive excitations, found by numerical solution of $D(\epsilon, k; \theta) = 0 $ (solid lines) and from~\eqref{eq:sol_strong} (dashed lines) for different values of the matching angle $\theta = 5^\circ, 15^\circ, 30^\circ, 90^\circ$ (the arrow shows the variation of the branch with increasing $\theta$).}
	\label{fig:gapped_spectrum}
\end{figure}

Solving the resultant quadratic equation with respect to $\bar{\epsilon} $, we find
\begin{equation}\label{eq:sol_strong}
 \bar{\epsilon}_{u,b}(\bar{k}) = \Delta(\pi/2) \cos(\theta/2) \pm
 			\sqrt{\bar{k}^2 + \Delta^2(\pi/2)},
\end{equation}
where
\begin{equation}\label{eq:gap_strong}
 \Delta(\pi/2) = \frac{K}{\sin(\theta/2)}
\end{equation}
is the mass of the low-energy excitations acquired due to the boundary induced coupling between helicities. Figure~\ref{fig:gapped_spectrum} presenting a comparison of these results with numerical solutions of $D(\epsilon, k; \theta) = 0 $ shows that the adopted approximation reproduces very well the spectrum of excitations near the extrema of the bands.

It must be emphasized that the acquired mass is the consequence of restricted geometry in the radial direction. With increasing the radius of the wire, the mass decreases and vanishes in the limit $R \to \infty $ corresponding to the well studied case of a WSM with a flat surface. At the same time, as follows from Eq.~\eqref{eq:gap_strong}, the acquired mass is not bounded from above and diverges as $\theta$ approaches $0$. In this limit, the spectrum consists of a single flat band at $\bar{\epsilon} = 0 $, while the second band escapes to infinity. The emergence of the flat band can be seen in the case $q_z = q_0 = 0$ directly from Eq.~\eqref{eq:D_strong_coupling}. Indeed, for $\theta = 0$, we have
\begin{equation}\label{eq:flat_expl}
 D(\epsilon, k; \theta = 0) = 2 \epsilon k J_{m}(Q r) J_{m+1}(Q r),
\end{equation}
which vanishes either at isolated points or when $\epsilon(k) \equiv 0$.

Finally, it should be noted that changing the sign of the gap when $\theta$ passes through zero corresponds to changing the character of the low energy band. It reaches minimum or maximum at $\bar{k} = 0$ and shifts upwards or downwards with increasing magnitude of $\theta$ for positive and negative $\theta$, respectively.

\subsection{General boundary conditions}
\label{sec:general_BC}

In view of Eq.~\eqref{eq:D_full_sep}, traversing the manifold of boundary conditions should yield a continuous transition to the strong coupling limit. As a result, the main spectral features found for the case $\Lambda = \pi/2$, in particular, opened gap in the spectrum of excitations massless when the helicities are fully decoupled, should preserve for the general case as well.

Using the same approach as in the previous section, we obtain an equation with respect to $\bar{\epsilon} $ with the solutions
\begin{equation}\label{eq:sol_gen}
 \bar{\epsilon}_{u,b}(\bar{k}) = \Delta(\Lambda) \frac{\sin(\delta)}{\sin{\Lambda}}  \pm
 			\sqrt{ \left[ \bar{k} - k_D(\Lambda)\right]^2 + \Delta^2(\Lambda)},
\end{equation}
where we have excluded $\theta$ using the relation $\beta_+ - \beta_- = \theta + \pi$ and denoting $\delta = (\beta_+ - \beta_-)/2$. The mass of low-energy excitations is 
\begin{equation}\label{eq:gap_gen}
 \Delta(\Lambda) = K \frac{\sin(\Lambda)}{\cos(\Lambda) \cos(\bar{\beta}) + \cos(\delta)}
\end{equation}
with $\bar{\beta} = (\beta_+ + \beta_-)/2$, and the position of the extrema of the bands is given by
\begin{equation}\label{eq:bot_gen}
 k_D(\Lambda) = \Delta(\Lambda) \cot(\Lambda) \sin(\bar{\beta}).
\end{equation}
Equation~\eqref{eq:gap_gen} defines a subset of boundary conditions producing the flat-band in the spectrum of WSM wires
\begin{equation}\label{eq:flat_band_gen}
 \cos(\Lambda) \cos(\bar{\beta}) + \cos(\delta) = 0.
\end{equation}

\begin{figure}[tb]
	\centering
	\includegraphics[width=4.5in]{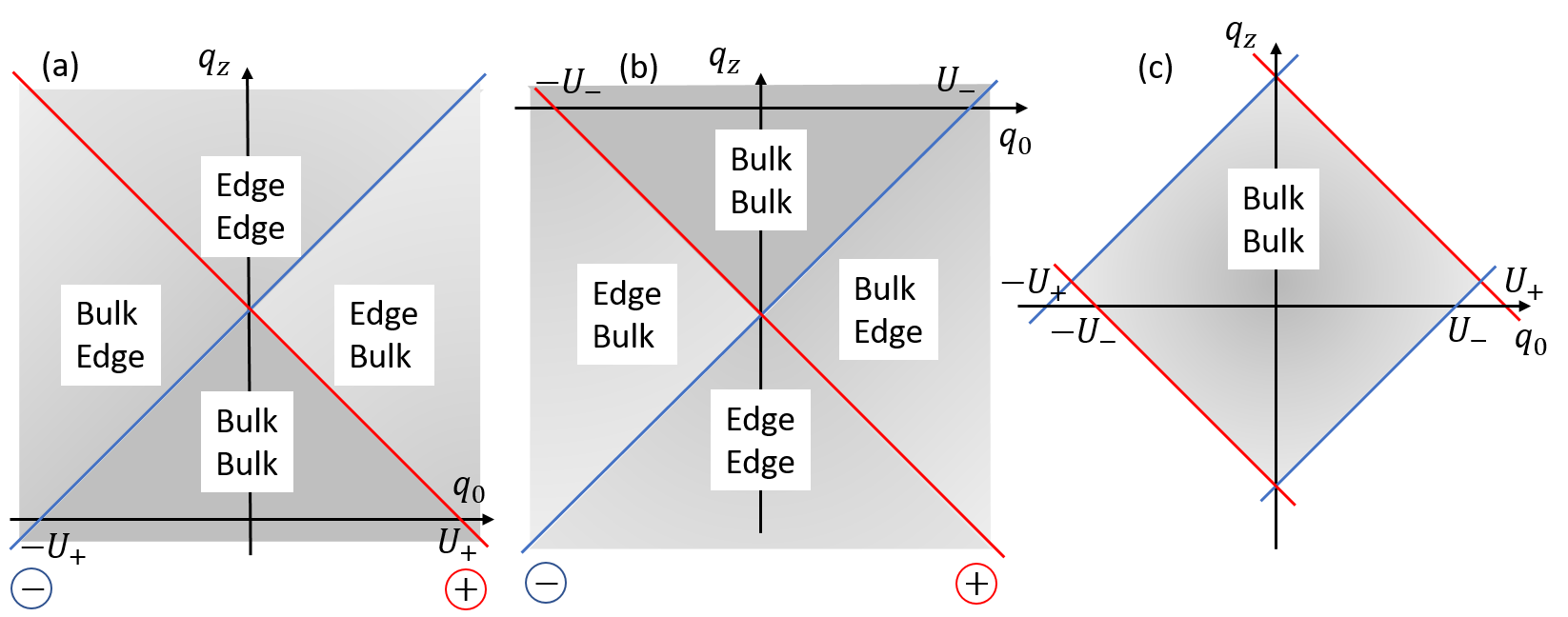}
	\caption{The phase diagram on the $(q_0, q_z)$-plane of the structure of states occupying the 
	upper (a) and the lower (b) branches (for $\theta > 0$) of the excitations gapped due to coupling
	between states with opposite helicities at the surface of the WSM wire. The transition lines corresponding to different helicities are identified by encircled $+$ and $-$. In specifications of the structure of the state (bulk or edge), the upper and lower lines are for $\xi = 1$ and $\xi = -1$, respectively. Panel (c) shows diagrams for different branches superimposed to indicate the region where all states have the same structure.}
	\label{fig:gapped_phases}
\end{figure}

Using the obtained results, one can show that states near the edges of the opened gap can be either bulk or surface depending on separation between the Weyl points. The character of the spatial variation of the state in the radial direction is determined by the sign of $Q_\xi^2 = \epsilon_\xi^2 - k_\xi^2 $. Substituting the solution obtained above, we obtain that the vicinity of the minimum of the upper (for $\delta > 0 $) branch is occupied by bulk modes, if
\begin{equation}\label{eq:upper_bulk}
 U_+ - (q_z + \xi q_0) > 0.
\end{equation}
In turn, the top of the lower brunch corresponds to bulk modes, if
\begin{equation}\label{eq:lower_bulk}
 U_- + (q_z + \xi q_0) > 0,
\end{equation}
where
\begin{equation}\label{eq:U_pm_par}
 U_\pm = \Delta(\Lambda) \left[ \frac{\sin(\delta)}{\sin(\Lambda)} \pm 1 \right].
\end{equation}
We present these results in Fig.~\ref{fig:gapped_phases} in the form of a phase diagram on the $(q_0,q_z) $-plane of separations between the Weyl points. The straight lines separate the regions where the extrema of the separated bands correspond to bulk and surface modes and are determined by the boundary conditions only. In Fig.~\ref{fig:gapped_phases}(c), we superimpose diagrams for the lower and upper bands to emphasize that for small separations $q_{0,z} $ the near-edge (non-relativistic) states are bulk modes. At the same time it should be noted that with increasing the radius of the wire, the middle region in Fig.~\ref{fig:gapped_phases}(c) shrinks relaxing the condition imposed on the separation of the Weyl points.

% \tobedone{Relation between helicities, conservation of helicitiy, states transmitting helicity, separation between the Weyl points along the axis of the wire, collapse of Fermi arcs}

% \section{The effect of separation of the Weyl points in the momentum space}

% \tobedone{Transversal separation: breaking the cylindircal symmetry}

% \begin{equation}\label{eq:M_transversal}
% \widehat{M}_\nu(\vect{v}, \theta, \chi) \to \widehat{M}_\nu(\vect{v}, \theta, \chi(\phi))
% \end{equation}
% where $\chi(\phi) = \chi + 2 q r_B \cos(\phi)$ and $q$ is the magnitude of the separation between the Weyl points in the momentum space.

% ... Separating contributions with different projections of the total angular momentum, we obtain instead of Eq.~\eqref{eq:Psi_BC}
% \begin{equation}\label{eq:Psi_ang_coupling}
%  \Psi^{(m)} = \sum_{m'} \widehat{M}^{(m-m')}_\nu \Psi^{(m')},
% \end{equation}
% where
% \begin{equation}\label{eq:M_momenta}
% \widehat{M}_\nu^{(m-m')} = \widehat{M}_\nu (\vect{v}) \cos(\theta) + J_{m- m'}(2 qr_B) \widehat{M}_\nu(\chi + \pi(m-m'))
% \end{equation}

\section{Conclusion}

We have studied the effect of restricting geometry on the spectrum of excitations in Weyl semimetals (WSM) using as a model a cylindrical WSM wire.

We have parametrized the full manifold of hard boundary conditions imposed by the requirement of vanishing amplitude of finding electron outside of the wire. The most general boundary conditions are described by two distributions of unit vectors tangent to the surface of the wire, and the distributions of a scalar coupling parameter and a relative phase between the states with different helicities.

We derive the general form of the dispersion equation relating energy and longitudinal momentum of electrons propagating along the wire. The coupling parameter describes a continuous transition between the limit of decoupled helicities when the boundary conditions have the form of pinned spin states at the surface of the wire, to the strong coupling case, where the rotational symmetry is restored at the surface but by the price of a direct relation between the orientations of the spins of electrons with different helicities.

We study main spectral features following from the derived dispersion equation. We show that in the limit of decoupled helicities, for each helicity, the spectrum consists of massive bands and a single massless band. At the same time, we show that the presence of massless bands is specific for systems with the boundary conditions supporting decoupled helicities. Since relaxing the spin state at the surface of the wire is achieved by coupling states with different helicities, this leads to lifting the degeneracy at the point of anti-crossing resonance with opening a gap in the spectrum and, thus, creating two massive modes. The acquired mass is a result of the confined geometry in the radial direction and vanishes in the limit of infinite radius of the wire (flat surface).

This demonstrates that restricting geometry may lead to strong modifications of the spectral properties of WSM and may lead to formation of new classes of excitations.

\section*{Acknowledgment}

The work was supported by the Air Force Office of Scientific Research (AFOSR) Grant No. FA9550-16-1-0363. 

\appendix

\section{Zero-mass modes}
\label{sec:zero_mass_app}

A straightforward substitution of $\epsilon_\xi = \xi k_\xi $ into~\eqref{eq:psi_radial} yields a trivial solution $\psi_\xi^{(j)}(r) \equiv 0$. Here, we derive a correct form of solutions of 
\begin{equation}\label{eq:reduced_eq}
 ({\epsilon} + \xi k \sigma_z) \psi(\rr_\perp) = \rmi \xi \bosi_\perp \cdot \nabla_\perp \psi(\rr_\perp)
\end{equation}
for $Q_\xi \equiv \sqrt{{\epsilon}^2 - k^2} = 0$. Here and below, we assume that both external potentials and separations between the Weyl points are absent, since they only lead to, generally speaking helicity dependent, shifts of the energy and longitudinal momentum.

We notice that $Q_\xi = 0$ corresponds to the case when ${\epsilon} + \xi k \sigma_z$ has a zero eigenvalue. This simplifies~\eqref{eq:reduced_eq} to a system of equations, which can be directly integrated using, for example, the characteristics method. This presents the solution in terms of analytical functions of $\tau_\pm = \rme^{\pm \rmi \phi} r $. Among the full set of solutions, we are the most interested in those induced by the series representation of the analytical functions, which establishes a connection with representation in terms of states with the definite projection of the angular momentum on the $z$-axis. Thus, taking into account the condition of regularity at the origin, it is convenient to introduce
\begin{equation}\label{eq:h_monomials}
 h_s^{(m)} = \left( \rme^{\rmi s \phi} r \right)^m,
\end{equation}
and to approach Eq.~\eqref{eq:reduced_eq} directly having in mind the relations
\begin{equation}\label{eq:h_rec}
 \left( \partial/\partial x \pm \rmi \partial/\partial y \right) h_s^{(m)} = m (1 \mp s) h_s^{(m - 1)}.
\end{equation}
This leads to a solution of Eq.~\eqref{eq:reduced_eq} in the form (up to a normalization factor)
\begin{equation}\label{eq:sol_red}
 \psi(r, \phi) = \mqty(a h_s^{(n)} \\ b h_s^{(m)}).
\end{equation}
One class of solutions corresponds to either $a$ or $b$ equal to zero, so that the spin is parallel to the $z$-axis. The spectrum of these excitations consists of two branches, ${\epsilon} = \pm k$, with $\psi_\xi^{(m)} = h_{-\xi}^{(m)} \ket{-\xi z} $ on ${\epsilon} = k$, and $\psi_\xi^{(m)} = h_\xi^{(m)} \ket{\xi z} $ on ${\epsilon} = -k$.

 Additionally, we find that for $s > 0$ there are solutions existing when ${\epsilon} = \xi k $ and $n = m-1$ with $a = \rmi m$ and $b = k$. For $s < 0$, the non-trivial solutions exist when ${\epsilon} = -\xi k $ and $n = m+1$ with $a = -k$ and $b = \rmi n$.

It should be emphasized that for both these classes of solutions, Weyl fermions with opposite helicities occupy the same spectral branch with opposite projections of the total angular momentum. This is, essentially, a consequence of the requirement of regularity at the origin. It should be noted in this regard, that in hollow wires, the power of monomials is no longer limited by non-negative values. This lifts restrictions on the sign of the angular momentum (cf modes of the electromagnetic field guided by hollow cables \cite{tamm_fundamentals_1979}).

\section{Parametrization of the family of local boundary conditions}
\label{sec:bc_app}

At the boundary of the wire, the wave function must satisfy a boundary condition $\Psi = \widehat{M}\Psi $ with such Hermitian $\widehat{M} $ that ensures vanishing current through the surface of the wire, $j_n = \matrixel{\Psi}{\vect{n}_B \cdot \boal}{\Psi} = 0 $ with $\vect{n}_B $ being the normal to the surface. Such boundary conditions in the context of the Dirac equation have been discussed in a number of publications (see, e.g., \cite{mccann_symmetry_2004,akhmerov_detection_2007,akhmerov_boundary_2008}). We, however, need $\widehat{M} $ represented in a way emphasizing coupling between particles with different helicities, which is slightly different from previously used representations.

The matrix $\widehat{M} $ must satisfy the anticommutation relation $\left\{ \widehat{M}, \vect{n}_B \cdot \boal \right\} = 0 $ and can be chosen unitary, so that $\widehat{M}^2 = \widehat{1}$. In the cylindrical coordinate system, in the rotated spinor frame with $\Psi = \exp \left( -\rmi \Sigma_z \phi/2 \right) \widetilde{\Psi} $, the radial component of $\boal$ turns into $\alpha_x$ and for $\widetilde{M} = \rme^{ \rmi \Sigma_z \phi/2} \widehat{M} \rme^{ -\rmi \Sigma_z \phi/2}$ we have
\begin{equation}\label{eq:M_anticom}
 \left\{ \widetilde{M}, \alpha_x \right\} = 0.
\end{equation}
The matrix $\widetilde{M} $ can be presented in the block form $\widetilde{M} = \mqty(M_{11} & M_{12} \\ M_{12}^\dagger & M_{22}) $, where $M_{ij} $ are $2\times 2$-matrices, which in the chiral representation satisfy $ \left\{ M_{ii}, \sigma_x \right\} = 0 $ and $ \left[ M_{12}, \sigma_x \right] = 0 $. Thus, for an arbitrary matrix $\widetilde{M} $ anticommuting with $\alpha_x$, we have $M_{ii} = \vect{m}_{ii} \cdot \bosi $, $M_{12} = m_{12}^{(0)} \sigma_0 + m_{12}^{(x)} \sigma_x $, where $\sigma_0$ is a $2\times2$ identity matrix,  vectors $\vect{m}_{ii} $ are tangent to the surface of the wire, and $m_{12}^{(0, x)}$ are arbitrary complex numbers.

Imposing the condition $\widehat{M}^2 = \widehat{1} $, we find that the parameters defined above are subjects of several constraints. For example, from equality of the diagonal blocks of $\widetilde{M}^2 $ to $\sigma_0$, it follows that $m_{ii}^2 + |m_{12}^{(0)}|^2 + |m_{12}^{(x)}|^2 = 1 $ and $ \mathrm{Re} \left[ m_{12}^{(0)} {m_{12}^{(1)}}^* \right] = 0$. They show that $M_{12} $ is proportional to a unitary matrix and thus the condition $M_{11}M_{12} + M_{12}M_{22} = 0 $ can be regarded as a relation between vectors $\vect{m}_{ii} $ and a condition imposed on $M_{12} $: it must leave vectors $\vect{m}_{ii} $ in the tangent plane. Making use of these observations, we can represent
\begin{equation}\label{eq:M_separated_app}
 \widetilde{M} = \widehat{M}_\|(\vect{v}_+, \vect{v}_- ) \cos{\Lambda}
 			+ \widehat{M}_\perp(\theta, \chi ) \sin{\Lambda},
\end{equation}
where $\Lambda $ is the helicity mixing angle, and
\begin{equation}\label{eq:M_constituents_app}
\begin{split}
 \widehat{M}_\|(\vect{v}_+, \vect{v}_- ) = & \mqty(\vect{v}_+ \cdot \bosi & 0 \\
 										 0 & \vect{v}_- \cdot \bosi), \\
 		\widehat{M}_\perp(\theta, \chi ) = & \mqty(0 & \rme^{\rmi \chi} R_x(\theta) \\ 
 											\rme^{-\rmi \chi} R_x^{-1}(\theta) & 0).
\end{split} 											
\end{equation}
Here $R_x(\theta) = \rme^{- \rmi \sigma_x \theta/2} $ is the rotation around the $x$-axis in the spin $1/2$ representation. Vectors $\vect{v}_{\pm} $ are unit vectors in the plane tangent to the surface of the wire and are related by $\vect{v}_+ = - R_x(\theta) \vect{v}_- $, which can be regarded as either a relation between $\vect{v}_\pm $ or a definition of $\theta$.

We consider two important particular cases: $\Lambda = 0$ and $\Lambda = \pi/2$. 

In the case $\Lambda = 0$, the states with opposite helicities are decoupled and satisfy their own boundary conditions: $\psi_\xi = \vect{v}_\xi \cdot \bosi \psi_\xi $. They correspond to fermion's spins taking definite directions determined by vectors $\vect{v}_{1,2} $, in other words the spins are pinned at the surface of the wire.

The second case, $\Lambda  = \pi/2 $, corresponds to strong coupling. An example of physical situation where such model appears naturally is the case when the electron is described by the Dirac equation with zero mass inside the wire and large mass outside. Similar models were widely applied for describing hard boundaries in reduced dimensionalities \cite{berry_m._v._neutrino_1987,erementchouk_optical_2015}. In this case, physical states are attenuated in the prohibited region due to the mass barrier and the boundary conditions are formulated as orthogonality to unphysical exponentially growing solutions. Solving the Dirac equation and taking the limit of infinite mass outside yields the boundary condition with $\widetilde{M} = \widehat{M}_\perp(\theta = \pi, \chi = - \pi/2)$. It should be noted, however, that the spin states inside the material should not necessarily correspond to the spin states in the prohibited region and, thus, a unitary transformation of spin states should be allowed at the boundary yielding the class of boundary conditions with arbitrary $\theta$. The important feature of this model is that it demonstrates that the boundary conditions can be determined by the environment only regardless of dynamical properties inside the wire, in particular, of the separation between the Weyl points.

% \bibliography{refs_wsm_wire}

\begin{thebibliography}{38}%
\makeatletter
\providecommand \@ifxundefined [1]{%
 \@ifx{#1\undefined}
}%
\providecommand \@ifnum [1]{%
 \ifnum #1\expandafter \@firstoftwo
 \else \expandafter \@secondoftwo
 \fi
}%
\providecommand \@ifx [1]{%
 \ifx #1\expandafter \@firstoftwo
 \else \expandafter \@secondoftwo
 \fi
}%
\providecommand \natexlab [1]{#1}%
\providecommand \enquote  [1]{``#1''}%
\providecommand \bibnamefont  [1]{#1}%
\providecommand \bibfnamefont [1]{#1}%
\providecommand \citenamefont [1]{#1}%
\providecommand \href@noop [0]{\@secondoftwo}%
\providecommand \href [0]{\begingroup \@sanitize@url \@href}%
\providecommand \@href[1]{\@@startlink{#1}\@@href}%
\providecommand \@@href[1]{\endgroup#1\@@endlink}%
\providecommand \@sanitize@url [0]{\catcode `\\12\catcode `\$12\catcode
  `\&12\catcode `\#12\catcode `\^12\catcode `\_12\catcode `\%12\relax}%
\providecommand \@@startlink[1]{}%
\providecommand \@@endlink[0]{}%
\providecommand \url  [0]{\begingroup\@sanitize@url \@url }%
\providecommand \@url [1]{\endgroup\@href {#1}{\urlprefix }}%
\providecommand \urlprefix  [0]{URL }%
\providecommand \Eprint [0]{\href }%
\providecommand \doibase [0]{http://dx.doi.org/}%
\providecommand \selectlanguage [0]{\@gobble}%
\providecommand \bibinfo  [0]{\@secondoftwo}%
\providecommand \bibfield  [0]{\@secondoftwo}%
\providecommand \translation [1]{[#1]}%
\providecommand \BibitemOpen [0]{}%
\providecommand \bibitemStop [0]{}%
\providecommand \bibitemNoStop [0]{.\EOS\space}%
\providecommand \EOS [0]{\spacefactor3000\relax}%
\providecommand \BibitemShut  [1]{\csname bibitem#1\endcsname}%
\let\auto@bib@innerbib\@empty
%</preamble>
\bibitem [{\citenamefont {Yang}\ \emph {et~al.}(2015)\citenamefont {Yang},
  \citenamefont {Liu}, \citenamefont {Sun}, \citenamefont {Peng}, \citenamefont
  {Yang}, \citenamefont {Zhang}, \citenamefont {Zhou}, \citenamefont {Zhang},
  \citenamefont {Guo}, \citenamefont {Rahn}, \citenamefont {Prabhakaran},
  \citenamefont {Hussain}, \citenamefont {Mo}, \citenamefont {Felser},
  \citenamefont {Yan},\ and\ \citenamefont {Chen}}]{yang_weyl_2015}%
  \BibitemOpen
  \bibfield  {author} {\bibinfo {author} {\bibfnamefont {L.~X.}\ \bibnamefont
  {Yang}}, \bibinfo {author} {\bibfnamefont {Z.~K.}\ \bibnamefont {Liu}},
  \bibinfo {author} {\bibfnamefont {Y.}~\bibnamefont {Sun}}, \bibinfo {author}
  {\bibfnamefont {H.}~\bibnamefont {Peng}}, \bibinfo {author} {\bibfnamefont
  {H.~F.}\ \bibnamefont {Yang}}, \bibinfo {author} {\bibfnamefont
  {T.}~\bibnamefont {Zhang}}, \bibinfo {author} {\bibfnamefont
  {B.}~\bibnamefont {Zhou}}, \bibinfo {author} {\bibfnamefont {Y.}~\bibnamefont
  {Zhang}}, \bibinfo {author} {\bibfnamefont {Y.~F.}\ \bibnamefont {Guo}},
  \bibinfo {author} {\bibfnamefont {M.}~\bibnamefont {Rahn}}, \bibinfo {author}
  {\bibfnamefont {D.}~\bibnamefont {Prabhakaran}}, \bibinfo {author}
  {\bibfnamefont {Z.}~\bibnamefont {Hussain}}, \bibinfo {author} {\bibfnamefont
  {S.-K.}\ \bibnamefont {Mo}}, \bibinfo {author} {\bibfnamefont
  {C.}~\bibnamefont {Felser}}, \bibinfo {author} {\bibfnamefont
  {B.}~\bibnamefont {Yan}}, \ and\ \bibinfo {author} {\bibfnamefont {Y.~L.}\
  \bibnamefont {Chen}},\ }\href {\doibase 10.1038/nphys3425} {\bibfield
  {journal} {\bibinfo  {journal} {Nat. Phys.}\ }\textbf {\bibinfo {volume}
  {11}},\ \bibinfo {pages} {728} (\bibinfo {year} {2015})}\BibitemShut
  {NoStop}%
\bibitem [{\citenamefont {Xu}\ \emph {et~al.}(2015{\natexlab{a}})\citenamefont
  {Xu}, \citenamefont {Liu}, \citenamefont {Kushwaha}, \citenamefont {Sankar},
  \citenamefont {Krizan}, \citenamefont {Belopolski}, \citenamefont {Neupane},
  \citenamefont {Bian}, \citenamefont {Alidoust}, \citenamefont {Chang},
  \citenamefont {Jeng}, \citenamefont {Huang}, \citenamefont {Tsai},
  \citenamefont {Lin}, \citenamefont {Shibayev}, \citenamefont {Chou},
  \citenamefont {Cava},\ and\ \citenamefont {Hasan}}]{xu_observation_2015}%
  \BibitemOpen
  \bibfield  {author} {\bibinfo {author} {\bibfnamefont {S.-Y.}\ \bibnamefont
  {Xu}}, \bibinfo {author} {\bibfnamefont {C.}~\bibnamefont {Liu}}, \bibinfo
  {author} {\bibfnamefont {S.~K.}\ \bibnamefont {Kushwaha}}, \bibinfo {author}
  {\bibfnamefont {R.}~\bibnamefont {Sankar}}, \bibinfo {author} {\bibfnamefont
  {J.~W.}\ \bibnamefont {Krizan}}, \bibinfo {author} {\bibfnamefont
  {I.}~\bibnamefont {Belopolski}}, \bibinfo {author} {\bibfnamefont
  {M.}~\bibnamefont {Neupane}}, \bibinfo {author} {\bibfnamefont
  {G.}~\bibnamefont {Bian}}, \bibinfo {author} {\bibfnamefont {N.}~\bibnamefont
  {Alidoust}}, \bibinfo {author} {\bibfnamefont {T.-R.}\ \bibnamefont {Chang}},
  \bibinfo {author} {\bibfnamefont {H.-T.}\ \bibnamefont {Jeng}}, \bibinfo
  {author} {\bibfnamefont {C.-Y.}\ \bibnamefont {Huang}}, \bibinfo {author}
  {\bibfnamefont {W.-F.}\ \bibnamefont {Tsai}}, \bibinfo {author}
  {\bibfnamefont {H.}~\bibnamefont {Lin}}, \bibinfo {author} {\bibfnamefont
  {P.~P.}\ \bibnamefont {Shibayev}}, \bibinfo {author} {\bibfnamefont {F.-C.}\
  \bibnamefont {Chou}}, \bibinfo {author} {\bibfnamefont {R.~J.}\ \bibnamefont
  {Cava}}, \ and\ \bibinfo {author} {\bibfnamefont {M.~Z.}\ \bibnamefont
  {Hasan}},\ }\href {\doibase 10.1126/science.1256742} {\bibfield  {journal}
  {\bibinfo  {journal} {Science}\ }\textbf {\bibinfo {volume} {347}},\ \bibinfo
  {pages} {294} (\bibinfo {year} {2015}{\natexlab{a}})}\BibitemShut {NoStop}%
\bibitem [{\citenamefont {Xu}\ \emph {et~al.}(2015{\natexlab{b}})\citenamefont
  {Xu}, \citenamefont {Belopolski}, \citenamefont {Alidoust}, \citenamefont
  {Neupane}, \citenamefont {Bian}, \citenamefont {Zhang}, \citenamefont
  {Sankar}, \citenamefont {Chang}, \citenamefont {Yuan}, \citenamefont {Lee},
  \citenamefont {Huang}, \citenamefont {Zheng}, \citenamefont {Ma},
  \citenamefont {Sanchez}, \citenamefont {Wang}, \citenamefont {Bansil},
  \citenamefont {Chou}, \citenamefont {Shibayev}, \citenamefont {Lin},
  \citenamefont {Jia},\ and\ \citenamefont {Hasan}}]{xu_discovery_2015}%
  \BibitemOpen
  \bibfield  {author} {\bibinfo {author} {\bibfnamefont {S.-Y.}\ \bibnamefont
  {Xu}}, \bibinfo {author} {\bibfnamefont {I.}~\bibnamefont {Belopolski}},
  \bibinfo {author} {\bibfnamefont {N.}~\bibnamefont {Alidoust}}, \bibinfo
  {author} {\bibfnamefont {M.}~\bibnamefont {Neupane}}, \bibinfo {author}
  {\bibfnamefont {G.}~\bibnamefont {Bian}}, \bibinfo {author} {\bibfnamefont
  {C.}~\bibnamefont {Zhang}}, \bibinfo {author} {\bibfnamefont
  {R.}~\bibnamefont {Sankar}}, \bibinfo {author} {\bibfnamefont
  {G.}~\bibnamefont {Chang}}, \bibinfo {author} {\bibfnamefont
  {Z.}~\bibnamefont {Yuan}}, \bibinfo {author} {\bibfnamefont {C.-C.}\
  \bibnamefont {Lee}}, \bibinfo {author} {\bibfnamefont {S.-M.}\ \bibnamefont
  {Huang}}, \bibinfo {author} {\bibfnamefont {H.}~\bibnamefont {Zheng}},
  \bibinfo {author} {\bibfnamefont {J.}~\bibnamefont {Ma}}, \bibinfo {author}
  {\bibfnamefont {D.~S.}\ \bibnamefont {Sanchez}}, \bibinfo {author}
  {\bibfnamefont {B.}~\bibnamefont {Wang}}, \bibinfo {author} {\bibfnamefont
  {A.}~\bibnamefont {Bansil}}, \bibinfo {author} {\bibfnamefont
  {F.}~\bibnamefont {Chou}}, \bibinfo {author} {\bibfnamefont {P.~P.}\
  \bibnamefont {Shibayev}}, \bibinfo {author} {\bibfnamefont {H.}~\bibnamefont
  {Lin}}, \bibinfo {author} {\bibfnamefont {S.}~\bibnamefont {Jia}}, \ and\
  \bibinfo {author} {\bibfnamefont {M.~Z.}\ \bibnamefont {Hasan}},\ }\href
  {\doibase 10.1126/science.aaa9297} {\bibfield  {journal} {\bibinfo  {journal}
  {Science}\ }\textbf {\bibinfo {volume} {349}},\ \bibinfo {pages} {613}
  (\bibinfo {year} {2015}{\natexlab{b}})}\BibitemShut {NoStop}%
\bibitem [{\citenamefont {Lv}\ \emph {et~al.}(2015{\natexlab{a}})\citenamefont
  {Lv}, \citenamefont {Xu}, \citenamefont {Weng}, \citenamefont {Ma},
  \citenamefont {Richard}, \citenamefont {Huang}, \citenamefont {Zhao},
  \citenamefont {Chen}, \citenamefont {Matt}, \citenamefont {Bisti},
  \citenamefont {Strocov}, \citenamefont {Mesot}, \citenamefont {Fang},
  \citenamefont {Dai}, \citenamefont {Qian}, \citenamefont {Shi},\ and\
  \citenamefont {Ding}}]{lv_observation_2015}%
  \BibitemOpen
  \bibfield  {author} {\bibinfo {author} {\bibfnamefont {B.~Q.}\ \bibnamefont
  {Lv}}, \bibinfo {author} {\bibfnamefont {N.}~\bibnamefont {Xu}}, \bibinfo
  {author} {\bibfnamefont {H.~M.}\ \bibnamefont {Weng}}, \bibinfo {author}
  {\bibfnamefont {J.~Z.}\ \bibnamefont {Ma}}, \bibinfo {author} {\bibfnamefont
  {P.}~\bibnamefont {Richard}}, \bibinfo {author} {\bibfnamefont {X.~C.}\
  \bibnamefont {Huang}}, \bibinfo {author} {\bibfnamefont {L.~X.}\ \bibnamefont
  {Zhao}}, \bibinfo {author} {\bibfnamefont {G.~F.}\ \bibnamefont {Chen}},
  \bibinfo {author} {\bibfnamefont {C.~E.}\ \bibnamefont {Matt}}, \bibinfo
  {author} {\bibfnamefont {F.}~\bibnamefont {Bisti}}, \bibinfo {author}
  {\bibfnamefont {V.~N.}\ \bibnamefont {Strocov}}, \bibinfo {author}
  {\bibfnamefont {J.}~\bibnamefont {Mesot}}, \bibinfo {author} {\bibfnamefont
  {Z.}~\bibnamefont {Fang}}, \bibinfo {author} {\bibfnamefont {X.}~\bibnamefont
  {Dai}}, \bibinfo {author} {\bibfnamefont {T.}~\bibnamefont {Qian}}, \bibinfo
  {author} {\bibfnamefont {M.}~\bibnamefont {Shi}}, \ and\ \bibinfo {author}
  {\bibfnamefont {H.}~\bibnamefont {Ding}},\ }\href {\doibase
  10.1038/nphys3426} {\bibfield  {journal} {\bibinfo  {journal} {Nat. Phys.}\
  }\textbf {\bibinfo {volume} {11}},\ \bibinfo {pages} {724} (\bibinfo {year}
  {2015}{\natexlab{a}})}\BibitemShut {NoStop}%
\bibitem [{\citenamefont {Lv}\ \emph {et~al.}(2015{\natexlab{b}})\citenamefont
  {Lv}, \citenamefont {Weng}, \citenamefont {Fu}, \citenamefont {Wang},
  \citenamefont {Miao}, \citenamefont {Ma}, \citenamefont {Richard},
  \citenamefont {Huang}, \citenamefont {Zhao}, \citenamefont {Chen},
  \citenamefont {Fang}, \citenamefont {Dai}, \citenamefont {Qian},\ and\
  \citenamefont {Ding}}]{lv_experimental_2015}%
  \BibitemOpen
  \bibfield  {author} {\bibinfo {author} {\bibfnamefont {B.~Q.}~\bibnamefont
  {Lv}}, \bibinfo {author} {\bibfnamefont {H.~M.}~\bibnamefont {Weng}}, \bibinfo
  {author} {\bibfnamefont {B.~B.}~\bibnamefont {Fu}}, \bibinfo {author}
  {\bibfnamefont {X.~P.}~\bibnamefont {Wang}}, \bibinfo {author} {\bibfnamefont
  {H.}~\bibnamefont {Miao}}, \bibinfo {author} {\bibfnamefont {J.}~\bibnamefont
  {Ma}}, \bibinfo {author} {\bibfnamefont {P.}~\bibnamefont {Richard}},
  \bibinfo {author} {\bibfnamefont {X.~C.}~\bibnamefont {Huang}}, \bibinfo
  {author} {\bibfnamefont {L.~X.}~\bibnamefont {Zhao}}, \bibinfo {author}
  {\bibfnamefont {G.~F.}~\bibnamefont {Chen}}, \bibinfo {author} {\bibfnamefont
  {Z.}~\bibnamefont {Fang}}, \bibinfo {author} {\bibfnamefont {X.}~\bibnamefont
  {Dai}}, \bibinfo {author} {\bibfnamefont {T.}~\bibnamefont {Qian}}, \ and\
  \bibinfo {author} {\bibfnamefont {H.}~\bibnamefont {Ding}},\ }\href {\doibase
  10.1103/PhysRevX.5.031013} {\bibfield  {journal} {\bibinfo  {journal} {Phys.
  Rev. X}\ }\textbf {\bibinfo {volume} {5}},\ \bibinfo {pages} {031013}
  (\bibinfo {year} {2015}{\natexlab{b}})}\BibitemShut {NoStop}%
\bibitem [{\citenamefont {Lu}\ \emph {et~al.}(2015)\citenamefont {Lu},
  \citenamefont {Wang}, \citenamefont {Ye}, \citenamefont {Ran}, \citenamefont
  {Fu}, \citenamefont {Joannopoulos},\ and\ \citenamefont
  {Soljačić}}]{lu_experimental_2015}%
  \BibitemOpen
  \bibfield  {author} {\bibinfo {author} {\bibfnamefont {L.}~\bibnamefont
  {Lu}}, \bibinfo {author} {\bibfnamefont {Z.}~\bibnamefont {Wang}}, \bibinfo
  {author} {\bibfnamefont {D.}~\bibnamefont {Ye}}, \bibinfo {author}
  {\bibfnamefont {L.}~\bibnamefont {Ran}}, \bibinfo {author} {\bibfnamefont
  {L.}~\bibnamefont {Fu}}, \bibinfo {author} {\bibfnamefont {J.~D.}\
  \bibnamefont {Joannopoulos}}, \ and\ \bibinfo {author} {\bibfnamefont
  {M.}~\bibnamefont {Soljačić}},\ }\href {\doibase 10.1126/science.aaa9273}
  {\bibfield  {journal} {\bibinfo  {journal} {Science}\ }\textbf {\bibinfo
  {volume} {349}},\ \bibinfo {pages} {622} (\bibinfo {year}
  {2015})}\BibitemShut {NoStop}%
\bibitem [{\citenamefont {Weyl}(1929)}]{weyl_elektron_1929}%
  \BibitemOpen
  \bibfield  {author} {\bibinfo {author} {\bibfnamefont {H.}~\bibnamefont
  {Weyl}},\ }\href {\doibase 10.1007/BF01339504} {\bibfield  {journal}
  {\bibinfo  {journal} {Z. Phys.}\ }\textbf {\bibinfo {volume} {56}},\ \bibinfo
  {pages} {330} (\bibinfo {year} {1929})}\BibitemShut {NoStop}%
\bibitem [{\citenamefont {Liu}\ and\ \citenamefont
  {Vanderbilt}(2014)}]{prev_1_Liu}%
  \BibitemOpen
  \bibfield  {author} {\bibinfo {author} {\bibfnamefont {J.}~\bibnamefont
  {Liu}}\ and\ \bibinfo {author} {\bibfnamefont {D.}~\bibnamefont
  {Vanderbilt}},\ }\href@noop {} {\bibfield  {journal} {\bibinfo  {journal}
  {Phys. Rev. B}\ }\textbf {\bibinfo {volume} {90}},\ \bibinfo {pages} {155316}
  (\bibinfo {year} {2014})}\BibitemShut {NoStop}%
\bibitem [{\citenamefont {Halasz}\ and\ \citenamefont
  {Balents}(2012)}]{prev_2_Halasz}%
  \BibitemOpen
  \bibfield  {author} {\bibinfo {author} {\bibfnamefont {G.~B.}\ \bibnamefont
  {Halasz}}\ and\ \bibinfo {author} {\bibfnamefont {L.}~\bibnamefont
  {Balents}},\ }\href@noop {} {\bibfield  {journal} {\bibinfo  {journal} {Phys.
  Rev. B}\ }\textbf {\bibinfo {volume} {85}},\ \bibinfo {pages} {035103}
  (\bibinfo {year} {2012})}\BibitemShut {NoStop}%
\bibitem [{\citenamefont {Shekhar}\ \emph {et~al.}(2015)\citenamefont
  {Shekhar}, \citenamefont {Nayak}, \citenamefont {Sun}, \citenamefont
  {Schmidt}, \citenamefont {Nicklas}, \citenamefont {Leermakers}, \citenamefont
  {Zeitler}, \citenamefont {Skourski}, \citenamefont {Wosnitza}, \citenamefont
  {Liu}, \citenamefont {Chen}, \citenamefont {Schnelle}, \citenamefont
  {Borrmann}, \citenamefont {Grin}, \citenamefont {Felser},\ and\ \citenamefont
  {Yan}}]{shekhar_extremely_2015}%
  \BibitemOpen
  \bibfield  {author} {\bibinfo {author} {\bibfnamefont {C.}~\bibnamefont
  {Shekhar}}, \bibinfo {author} {\bibfnamefont {A.~K.}\ \bibnamefont {Nayak}},
  \bibinfo {author} {\bibfnamefont {Y.}~\bibnamefont {Sun}}, \bibinfo {author}
  {\bibfnamefont {M.}~\bibnamefont {Schmidt}}, \bibinfo {author} {\bibfnamefont
  {M.}~\bibnamefont {Nicklas}}, \bibinfo {author} {\bibfnamefont
  {I.}~\bibnamefont {Leermakers}}, \bibinfo {author} {\bibfnamefont
  {U.}~\bibnamefont {Zeitler}}, \bibinfo {author} {\bibfnamefont
  {Y.}~\bibnamefont {Skourski}}, \bibinfo {author} {\bibfnamefont
  {J.}~\bibnamefont {Wosnitza}}, \bibinfo {author} {\bibfnamefont
  {Z.}~\bibnamefont {Liu}}, \bibinfo {author} {\bibfnamefont {Y.}~\bibnamefont
  {Chen}}, \bibinfo {author} {\bibfnamefont {W.}~\bibnamefont {Schnelle}},
  \bibinfo {author} {\bibfnamefont {H.}~\bibnamefont {Borrmann}}, \bibinfo
  {author} {\bibfnamefont {Y.}~\bibnamefont {Grin}}, \bibinfo {author}
  {\bibfnamefont {C.}~\bibnamefont {Felser}}, \ and\ \bibinfo {author}
  {\bibfnamefont {B.}~\bibnamefont {Yan}},\ }\href {\doibase 10.1038/nphys3372}
  {\bibfield  {journal} {\bibinfo  {journal} {Nat. Phys.}\ }\textbf {\bibinfo
  {volume} {11}},\ \bibinfo {pages} {645} (\bibinfo {year} {2015})}\BibitemShut
  {NoStop}%
\bibitem [{\citenamefont {Wang}\ \emph {et~al.}(2016)\citenamefont {Wang},
  \citenamefont {Zheng}, \citenamefont {Shen}, \citenamefont {Lu},
  \citenamefont {Fang}, \citenamefont {Sheng}, \citenamefont {Zhou},
  \citenamefont {Yang}, \citenamefont {Li}, \citenamefont {Feng},\ and\
  \citenamefont {Xu}}]{wang_helicity-protected_2016}%
  \BibitemOpen
  \bibfield  {author} {\bibinfo {author} {\bibfnamefont {Z.}~\bibnamefont
  {Wang}}, \bibinfo {author} {\bibfnamefont {Y.}~\bibnamefont {Zheng}},
  \bibinfo {author} {\bibfnamefont {Z.}~\bibnamefont {Shen}}, \bibinfo {author}
  {\bibfnamefont {Y.}~\bibnamefont {Lu}}, \bibinfo {author} {\bibfnamefont
  {H.}~\bibnamefont {Fang}}, \bibinfo {author} {\bibfnamefont {F.}~\bibnamefont
  {Sheng}}, \bibinfo {author} {\bibfnamefont {Y.}~\bibnamefont {Zhou}},
  \bibinfo {author} {\bibfnamefont {X.}~\bibnamefont {Yang}}, \bibinfo {author}
  {\bibfnamefont {Y.}~\bibnamefont {Li}}, \bibinfo {author} {\bibfnamefont
  {C.}~\bibnamefont {Feng}}, \ and\ \bibinfo {author} {\bibfnamefont {Z.-A.}\
  \bibnamefont {Xu}},\ }\href {\doibase 10.1103/PhysRevB.93.121112} {\bibfield
  {journal} {\bibinfo  {journal} {Phys. Rev. B}\ }\textbf {\bibinfo {volume}
  {93}},\ \bibinfo {pages} {121112} (\bibinfo {year} {2016})}\BibitemShut
  {NoStop}%
\bibitem [{\citenamefont {Landsteiner}(2014)}]{landsteiner_anomalous_2014}%
  \BibitemOpen
  \bibfield  {author} {\bibinfo {author} {\bibfnamefont {K.}~\bibnamefont
  {Landsteiner}},\ }\href {\doibase 10.1103/PhysRevB.89.075124} {\bibfield
  {journal} {\bibinfo  {journal} {Phys. Rev. B}\ }\textbf {\bibinfo {volume}
  {89}},\ \bibinfo {pages} {075124} (\bibinfo {year} {2014})}\BibitemShut
  {NoStop}%
\bibitem [{\citenamefont {Landsteiner}(2016)}]{landsteiner_notes_2016}%
  \BibitemOpen
  \bibfield  {author} {\bibinfo {author} {\bibfnamefont {K.}~\bibnamefont
  {Landsteiner}},\ }\href {\doibase 10.5506/APhysPolB.47.2617} {\bibfield
  {journal} {\bibinfo  {journal} {Acta Physica Polonica B}\ }\textbf {\bibinfo
  {volume} {47}},\ \bibinfo {pages} {2617} (\bibinfo {year}
  {2016})}\BibitemShut {NoStop}%
\bibitem [{\citenamefont {Gorbar}\ \emph
  {et~al.}(2017{\natexlab{a}})\citenamefont {Gorbar}, \citenamefont {Miransky},
  \citenamefont {Shovkovy},\ and\ \citenamefont
  {Sukhachov}}]{gorbar_second-order_2017}%
  \BibitemOpen
  \bibfield  {author} {\bibinfo {author} {\bibfnamefont {E.~V.}\ \bibnamefont
  {Gorbar}}, \bibinfo {author} {\bibfnamefont {V.~A.}\ \bibnamefont
  {Miransky}}, \bibinfo {author} {\bibfnamefont {I.~A.}\ \bibnamefont
  {Shovkovy}}, \ and\ \bibinfo {author} {\bibfnamefont {P.~O.}\ \bibnamefont
  {Sukhachov}},\ }\href {\doibase 10.1103/PhysRevB.95.205141} {\bibfield
  {journal} {\bibinfo  {journal} {Phys. Rev. B}\ }\textbf {\bibinfo {volume}
  {95}},\ \bibinfo {pages} {205141} (\bibinfo {year}
  {2017}{\natexlab{a}})}\BibitemShut {NoStop}%
\bibitem [{\citenamefont {Qiu}\ \emph {et~al.}(2017)\citenamefont {Qiu},
  \citenamefont {Cao},\ and\ \citenamefont {Huang}}]{qiu_electrodynamics_2017}%
  \BibitemOpen
  \bibfield  {author} {\bibinfo {author} {\bibfnamefont {Z.}~\bibnamefont
  {Qiu}}, \bibinfo {author} {\bibfnamefont {G.}~\bibnamefont {Cao}}, \ and\
  \bibinfo {author} {\bibfnamefont {X.-G.}\ \bibnamefont {Huang}},\ }\href
  {\doibase 10.1103/PhysRevD.95.036002} {\bibfield  {journal} {\bibinfo
  {journal} {Phys. Rev. D}\ }\textbf {\bibinfo {volume} {95}},\ \bibinfo
  {pages} {036002} (\bibinfo {year} {2017})}\BibitemShut {NoStop}%
\bibitem [{\citenamefont {Cortijo}\ \emph {et~al.}(2016)\citenamefont
  {Cortijo}, \citenamefont {Kharzeev}, \citenamefont {Landsteiner},\ and\
  \citenamefont {Vozmediano}}]{cortijo_strain-induced_2016}%
  \BibitemOpen
  \bibfield  {author} {\bibinfo {author} {\bibfnamefont {A.}~\bibnamefont
  {Cortijo}}, \bibinfo {author} {\bibfnamefont {D.}~\bibnamefont {Kharzeev}},
  \bibinfo {author} {\bibfnamefont {K.}~\bibnamefont {Landsteiner}}, \ and\
  \bibinfo {author} {\bibfnamefont {M.~A.~H.}\ \bibnamefont {Vozmediano}},\
  }\href {\doibase 10.1103/PhysRevB.94.241405} {\bibfield  {journal} {\bibinfo
  {journal} {Phys. Rev. B}\ }\textbf {\bibinfo {volume} {94}},\ \bibinfo
  {pages} {241405} (\bibinfo {year} {2016})}\BibitemShut {NoStop}%
\bibitem [{\citenamefont {Grushin}\ \emph {et~al.}(2016)\citenamefont
  {Grushin}, \citenamefont {Venderbos}, \citenamefont {Vishwanath},\ and\
  \citenamefont {Ilan}}]{grushin_inhomogeneous_2016}%
  \BibitemOpen
  \bibfield  {author} {\bibinfo {author} {\bibfnamefont {A.~G.}\ \bibnamefont
  {Grushin}}, \bibinfo {author} {\bibfnamefont {J.~W.~F.}\ \bibnamefont
  {Venderbos}}, \bibinfo {author} {\bibfnamefont {A.}~\bibnamefont
  {Vishwanath}}, \ and\ \bibinfo {author} {\bibfnamefont {R.}~\bibnamefont
  {Ilan}},\ }\href {\doibase 10.1103/PhysRevX.6.041046} {\bibfield  {journal}
  {\bibinfo  {journal} {Phys. Rev. X}\ }\textbf {\bibinfo {volume} {6}},\
  \bibinfo {pages} {041046} (\bibinfo {year} {2016})}\BibitemShut {NoStop}%
\bibitem [{\citenamefont {Gorbar}\ \emph
  {et~al.}(2017{\natexlab{b}})\citenamefont {Gorbar}, \citenamefont {Miransky},
  \citenamefont {Shovkovy},\ and\ \citenamefont
  {Sukhachov}}]{gorbar_consistent_2017}%
  \BibitemOpen
  \bibfield  {author} {\bibinfo {author} {\bibfnamefont {E.~V.}\ \bibnamefont
  {Gorbar}}, \bibinfo {author} {\bibfnamefont {V.~A.}\ \bibnamefont
  {Miransky}}, \bibinfo {author} {\bibfnamefont {I.~A.}\ \bibnamefont
  {Shovkovy}}, \ and\ \bibinfo {author} {\bibfnamefont {P.~O.}\ \bibnamefont
  {Sukhachov}},\ }\href {\doibase 10.1103/PhysRevLett.118.127601} {\bibfield
  {journal} {\bibinfo  {journal} {Phys. Rev. Lett.}\ }\textbf {\bibinfo
  {volume} {118}},\ \bibinfo {pages} {127601} (\bibinfo {year}
  {2017}{\natexlab{b}})}\BibitemShut {NoStop}%
\bibitem [{\citenamefont {Zubkov}(2015)}]{zubkov_emergent_2015}%
  \BibitemOpen
  \bibfield  {author} {\bibinfo {author} {\bibfnamefont {M.}~\bibnamefont
  {Zubkov}},\ }\href {\doibase 10.1016/j.aop.2015.05.032} {\bibfield  {journal}
  {\bibinfo  {journal} {Ann. Phys.}\ }\textbf {\bibinfo {volume} {360}},\
  \bibinfo {pages} {655} (\bibinfo {year} {2015})}\BibitemShut {NoStop}%
\bibitem [{\citenamefont {Hidaka}\ \emph {et~al.}(2017)\citenamefont {Hidaka},
  \citenamefont {Pu},\ and\ \citenamefont {Yang}}]{hidaka_relativistic_2017}%
  \BibitemOpen
  \bibfield  {author} {\bibinfo {author} {\bibfnamefont {Y.}~\bibnamefont
  {Hidaka}}, \bibinfo {author} {\bibfnamefont {S.}~\bibnamefont {Pu}}, \ and\
  \bibinfo {author} {\bibfnamefont {D.-L.}\ \bibnamefont {Yang}},\ }\href
  {\doibase 10.1103/PhysRevD.95.091901} {\bibfield  {journal} {\bibinfo
  {journal} {Phys. Rev. D}\ }\textbf {\bibinfo {volume} {95}},\ \bibinfo
  {pages} {091901} (\bibinfo {year} {2017})}\BibitemShut {NoStop}%
\bibitem [{\citenamefont {Hosur}\ and\ \citenamefont
  {Qi}(2013)}]{hosur_recent_2013}%
  \BibitemOpen
  \bibfield  {author} {\bibinfo {author} {\bibfnamefont {P.}~\bibnamefont
  {Hosur}}\ and\ \bibinfo {author} {\bibfnamefont {X.}~\bibnamefont {Qi}},\
  }\href {\doibase 10.1016/j.crhy.2013.10.010} {\bibfield  {journal} {\bibinfo
  {journal} {Compt. Rend. Phys.}\ }\textbf {\bibinfo {volume} {14}},\ \bibinfo
  {pages} {857} (\bibinfo {year} {2013})}\BibitemShut {NoStop}%
\bibitem [{\citenamefont {Bai}\ \emph {et~al.}(2016{\natexlab{a}})\citenamefont
  {Bai}, \citenamefont {Yang},\ and\ \citenamefont
  {Wei}}]{bai_wavevector_2016}%
  \BibitemOpen
  \bibfield  {author} {\bibinfo {author} {\bibfnamefont {C.}~\bibnamefont
  {Bai}}, \bibinfo {author} {\bibfnamefont {Y.}~\bibnamefont {Yang}}, \ and\
  \bibinfo {author} {\bibfnamefont {K.-W.}\ \bibnamefont {Wei}},\ }\href
  {\doibase 10.1016/j.physleta.2015.12.002} {\bibfield  {journal} {\bibinfo
  {journal} {Phys. Lett. A}\ }\textbf {\bibinfo {volume} {380}},\ \bibinfo
  {pages} {764} (\bibinfo {year} {2016}{\natexlab{a}})}\BibitemShut {NoStop}%
\bibitem [{\citenamefont {Bai}\ \emph {et~al.}(2016{\natexlab{b}})\citenamefont
  {Bai}, \citenamefont {Yang},\ and\ \citenamefont {Chang}}]{bai_chiral_2016}%
  \BibitemOpen
  \bibfield  {author} {\bibinfo {author} {\bibfnamefont {C.}~\bibnamefont
  {Bai}}, \bibinfo {author} {\bibfnamefont {Y.}~\bibnamefont {Yang}}, \ and\
  \bibinfo {author} {\bibfnamefont {K.}~\bibnamefont {Chang}},\ }\href
  {\doibase 10.1038/srep21283} {\bibfield  {journal} {\bibinfo  {journal} {Sci.
  Rep.}\ }\textbf {\bibinfo {volume} {6}},\ \bibinfo {pages} {21283} (\bibinfo
  {year} {2016}{\natexlab{b}})}\BibitemShut {NoStop}%
\bibitem [{\citenamefont {Erementchouk}\ and\ \citenamefont
  {Mazumder}(2017)}]{erementchouk_helicity-dependent_2017}%
  \BibitemOpen
  \bibfield  {author} {\bibinfo {author} {\bibfnamefont {M.}~\bibnamefont
  {Erementchouk}}\ and\ \bibinfo {author} {\bibfnamefont {P.}~\bibnamefont
  {Mazumder}},\ }\href {\doibase 10.1016/j.physleta.2017.06.055} {\bibfield
  {journal} {\bibinfo  {journal} {Phys. Lett. A}\ }\textbf {\bibinfo {volume}
  {381}},\ \bibinfo {pages} {2866} (\bibinfo {year} {2017})}\BibitemShut
  {NoStop}%
\bibitem [{\citenamefont {Volkov}\ and\ \citenamefont
  {Enaldiev}(2016)}]{volkov_surface_2016}%
  \BibitemOpen
  \bibfield  {author} {\bibinfo {author} {\bibfnamefont {V.~A.}\ \bibnamefont
  {Volkov}}\ and\ \bibinfo {author} {\bibfnamefont {V.~V.}\ \bibnamefont
  {Enaldiev}},\ }\href {\doibase 10.1134/S1063776116030213} {\bibfield
  {journal} {\bibinfo  {journal} {J. Exp. Theor. Phys.}\ }\textbf {\bibinfo
  {volume} {122}},\ \bibinfo {pages} {608} (\bibinfo {year}
  {2016})}\BibitemShut {NoStop}%
\bibitem [{\citenamefont {Devizorova}\ and\ \citenamefont
  {Volkov}(2017)}]{devizorova_fermi_2017}%
  \BibitemOpen
  \bibfield  {author} {\bibinfo {author} {\bibfnamefont {Z.~A.}\ \bibnamefont
  {Devizorova}}\ and\ \bibinfo {author} {\bibfnamefont {V.~A.}\ \bibnamefont
  {Volkov}},\ }\href {\doibase 10.1103/PhysRevB.95.081302} {\bibfield
  {journal} {\bibinfo  {journal} {Phys. Rev. B}\ }\textbf {\bibinfo {volume}
  {95}},\ \bibinfo {pages} {081302} (\bibinfo {year} {2017})}\BibitemShut
  {NoStop}%
\bibitem [{\citenamefont {Li}\ and\ \citenamefont
  {Andreev}(2015)}]{li_spiraling_2015}%
  \BibitemOpen
  \bibfield  {author} {\bibinfo {author} {\bibfnamefont {S.}~\bibnamefont
  {Li}}\ and\ \bibinfo {author} {\bibfnamefont {A.~V.}\ \bibnamefont
  {Andreev}},\ }\href {\doibase 10.1103/PhysRevB.92.201107} {\bibfield
  {journal} {\bibinfo  {journal} {Phys. Rev. B}\ }\textbf {\bibinfo {volume}
  {92}},\ \bibinfo {pages} {201107} (\bibinfo {year} {2015})}\BibitemShut
  {NoStop}%
\bibitem [{\citenamefont {Hashimoto}\ \emph {et~al.}(2017)\citenamefont
  {Hashimoto}, \citenamefont {Kimura},\ and\ \citenamefont
  {Wu}}]{hashimoto_boundary_2017}%
  \BibitemOpen
  \bibfield  {author} {\bibinfo {author} {\bibfnamefont {K.}~\bibnamefont
  {Hashimoto}}, \bibinfo {author} {\bibfnamefont {T.}~\bibnamefont {Kimura}}, \
  and\ \bibinfo {author} {\bibfnamefont {X.}~\bibnamefont {Wu}},\ }\href
  {\doibase 10.1093/ptep/ptx053} {\bibfield  {journal} {\bibinfo  {journal}
  {Prog. Theor. Exper. Phys.}\ }\textbf {\bibinfo {volume} {2017}},\ \bibinfo
  {pages} {053101} (\bibinfo {year} {2017})}\BibitemShut {NoStop}%
\bibitem [{\citenamefont {Katsnelson}\ \emph {et~al.}(2006)\citenamefont
  {Katsnelson}, \citenamefont {Novoselov},\ and\ \citenamefont
  {Geim}}]{katsnelson_chiral_2006}%
  \BibitemOpen
  \bibfield  {author} {\bibinfo {author} {\bibfnamefont {M.~I.}\ \bibnamefont
  {Katsnelson}}, \bibinfo {author} {\bibfnamefont {K.~S.}\ \bibnamefont
  {Novoselov}}, \ and\ \bibinfo {author} {\bibfnamefont {A.~K.}\ \bibnamefont
  {Geim}},\ }\href {\doibase 10.1038/nphys384} {\bibfield  {journal} {\bibinfo
  {journal} {Nat. Phys.}\ }\textbf {\bibinfo {volume} {2}},\ \bibinfo {pages}
  {620} (\bibinfo {year} {2006})}\BibitemShut {NoStop}%
\bibitem [{\citenamefont {Tudorovskiy}\ \emph {et~al.}(2012)\citenamefont
  {Tudorovskiy}, \citenamefont {Reijnders},\ and\ \citenamefont
  {Katsnelson}}]{tudorovskiy_chiral_2012}%
  \BibitemOpen
  \bibfield  {author} {\bibinfo {author} {\bibfnamefont {T.}~\bibnamefont
  {Tudorovskiy}}, \bibinfo {author} {\bibfnamefont {K.~J.~A.}\ \bibnamefont
  {Reijnders}}, \ and\ \bibinfo {author} {\bibfnamefont {M.~I.}\ \bibnamefont
  {Katsnelson}},\ }\href {\doibase 10.1088/0031-8949/2012/T146/014010}
  {\bibfield  {journal} {\bibinfo  {journal} {Phys. Scr.}\ }\textbf {\bibinfo
  {volume} {2012}},\ \bibinfo {pages} {014010} (\bibinfo {year}
  {2012})}\BibitemShut {NoStop}%
\bibitem [{\citenamefont {Reijnders}\ \emph {et~al.}(2013)\citenamefont
  {Reijnders}, \citenamefont {Tudorovskiy},\ and\ \citenamefont
  {Katsnelson}}]{reijnders_semiclassical_2013}%
  \BibitemOpen
  \bibfield  {author} {\bibinfo {author} {\bibfnamefont {K.}~\bibnamefont
  {Reijnders}}, \bibinfo {author} {\bibfnamefont {T.}~\bibnamefont
  {Tudorovskiy}}, \ and\ \bibinfo {author} {\bibfnamefont {M.}~\bibnamefont
  {Katsnelson}},\ }\href {\doibase 10.1016/j.aop.2013.03.001} {\bibfield
  {journal} {\bibinfo  {journal} {Ann. Phys. (NY)}\ }\textbf {\bibinfo {volume}
  {333}},\ \bibinfo {pages} {155} (\bibinfo {year} {2013})}\BibitemShut
  {NoStop}%
\bibitem [{\citenamefont {Kleptsyn}\ \emph {et~al.}(2015)\citenamefont
  {Kleptsyn}, \citenamefont {Okunev}, \citenamefont {Schurov}, \citenamefont
  {Zubov},\ and\ \citenamefont {Katsnelson}}]{kleptsyn_chiral_2015}%
  \BibitemOpen
  \bibfield  {author} {\bibinfo {author} {\bibfnamefont {V.}~\bibnamefont
  {Kleptsyn}}, \bibinfo {author} {\bibfnamefont {A.}~\bibnamefont {Okunev}},
  \bibinfo {author} {\bibfnamefont {I.}~\bibnamefont {Schurov}}, \bibinfo
  {author} {\bibfnamefont {D.}~\bibnamefont {Zubov}}, \ and\ \bibinfo {author}
  {\bibfnamefont {M.~I.}\ \bibnamefont {Katsnelson}},\ }\href {\doibase
  10.1103/PhysRevB.92.165407} {\bibfield  {journal} {\bibinfo  {journal} {Phys.
  Rev. B}\ }\textbf {\bibinfo {volume} {92}},\ \bibinfo {pages} {165407}
  (\bibinfo {year} {2015})}\BibitemShut {NoStop}%
\bibitem [{\citenamefont {McCann}\ and\ \citenamefont
  {Fal'ko}(2004)}]{mccann_symmetry_2004}%
  \BibitemOpen
  \bibfield  {author} {\bibinfo {author} {\bibfnamefont {E.}~\bibnamefont
  {McCann}}\ and\ \bibinfo {author} {\bibfnamefont {V.~I.}\ \bibnamefont
  {Fal'ko}},\ }\href {\doibase 10.1088/0953-8984/16/13/016} {\bibfield
  {journal} {\bibinfo  {journal} {J. Phys.: Cond. Matter}\ }\textbf {\bibinfo
  {volume} {16}},\ \bibinfo {pages} {2371} (\bibinfo {year}
  {2004})}\BibitemShut {NoStop}%
\bibitem [{\citenamefont {Akhmerov}\ and\ \citenamefont
  {Beenakker}(2007)}]{akhmerov_detection_2007}%
  \BibitemOpen
  \bibfield  {author} {\bibinfo {author} {\bibfnamefont {A.~R.}\ \bibnamefont
  {Akhmerov}}\ and\ \bibinfo {author} {\bibfnamefont {C.~W.~J.}\ \bibnamefont
  {Beenakker}},\ }\href {\doibase 10.1103/PhysRevLett.98.157003} {\bibfield
  {journal} {\bibinfo  {journal} {Phys. Rev. Lett.}\ }\textbf {\bibinfo
  {volume} {98}},\ \bibinfo {pages} {157003} (\bibinfo {year}
  {2007})}\BibitemShut {NoStop}%
\bibitem [{\citenamefont {Akhmerov}\ and\ \citenamefont
  {Beenakker}(2008)}]{akhmerov_boundary_2008}%
  \BibitemOpen
  \bibfield  {author} {\bibinfo {author} {\bibfnamefont {A.~R.}\ \bibnamefont
  {Akhmerov}}\ and\ \bibinfo {author} {\bibfnamefont {C.~W.~J.}\ \bibnamefont
  {Beenakker}},\ }\href {\doibase 10.1103/PhysRevB.77.085423} {\bibfield
  {journal} {\bibinfo  {journal} {Phys. Rev. B}\ }\textbf {\bibinfo {volume}
  {77}},\ \bibinfo {pages} {085423} (\bibinfo {year} {2008})}\BibitemShut
  {NoStop}%
\bibitem [{\citenamefont {Tamm}(1979)}]{tamm_fundamentals_1979}%
  \BibitemOpen
  \bibfield  {author} {\bibinfo {author} {\bibfnamefont {I.~E.}\ \bibnamefont
  {Tamm}},\ }\href@noop {} {\emph {\bibinfo {title} {Fundamentals of the theory
  of electricity}}}\ (\bibinfo  {publisher} {Mir Publishers},\ \bibinfo
  {address} {Moscow},\ \bibinfo {year} {1979})\BibitemShut {NoStop}%
\bibitem [{\citenamefont {{Berry, M. V.}}\ and\ \citenamefont {{Mondragon, R.
  J.}}(1987)}]{berry_m._v._neutrino_1987}%
  \BibitemOpen
  \bibfield  {author} {\bibinfo {author} {\bibnamefont {{Berry, M. V.}}}\ and\
  \bibinfo {author} {\bibnamefont {{Mondragon, R. J.}}},\ }\href@noop {}
  {\bibfield  {journal} {\bibinfo  {journal} {Proc. R. Soc. Lond. A}\ }\textbf
  {\bibinfo {volume} {412}},\ \bibinfo {pages} {53} (\bibinfo {year}
  {1987})}\BibitemShut {NoStop}%
\bibitem [{\citenamefont {Erementchouk}\ \emph {et~al.}(2015)\citenamefont
  {Erementchouk}, \citenamefont {Khan},\ and\ \citenamefont
  {Leuenberger}}]{erementchouk_optical_2015}%
  \BibitemOpen
  \bibfield  {author} {\bibinfo {author} {\bibfnamefont {M.}~\bibnamefont
  {Erementchouk}}, \bibinfo {author} {\bibfnamefont {M.~A.}\ \bibnamefont
  {Khan}}, \ and\ \bibinfo {author} {\bibfnamefont {M.~N.}\ \bibnamefont
  {Leuenberger}},\ }\href {\doibase 10.1103/PhysRevB.92.121401} {\bibfield
  {journal} {\bibinfo  {journal} {Phys. Rev. B}\ }\textbf {\bibinfo {volume}
  {92}},\ \bibinfo {pages} {121401(R)} (\bibinfo {year} {2015})}\BibitemShut
  {NoStop}%
\end{thebibliography}
% \bibliographystyle{apsrev4-1}


\end{document}